\documentclass[aps,twocolumn,floats,superscriptaddress,prd,nofootinbib]{revtex4-1}
\newlength{\picwidth}
\usepackage{graphicx,epsfig,amsmath,amssymb,amsfonts,enumerate,bm}
\usepackage{color}
\usepackage{hyperref}

\definecolor{darkgreen}{cmyk}{0.85,0.2,1.00,0.2}

\newcommand{\lcdm}{$\Lambda$CDM}

\begin{document}
\def\be{\begin{equation}}
\def\ee{\end{equation}}
\def\ba{\begin{eqnarray}}
\def\ea{\end{eqnarray}}

\title{Testing dark energy paradigms with weak gravitational lensing}

\author{R. Ali Vanderveld}
\affiliation{Kavli Institute for Cosmological Physics, Enrico Fermi Institute, University of Chicago, Chicago, IL 60637}
\author{Michael J. Mortonson}
\affiliation{Center for Cosmology and AstroParticle Physics, The Ohio State University, Columbus, OH 43210}
\author{Wayne Hu}
\affiliation{Kavli Institute for Cosmological Physics, Enrico Fermi Institute, University of Chicago, Chicago, IL 60637}
\affiliation{Department of Astronomy and Astrophysics, University of Chicago, Chicago, IL 60637}
\author{Tim Eifler}
\affiliation{Center for Cosmology and AstroParticle Physics, The Ohio State University, Columbus, OH 43210}
\date{\today}

\begin{abstract}

Any theory invoked to explain cosmic acceleration predicts consistency relations between the expansion history, structure growth, and all related observables. Currently there exist high-quality measurements of the expansion history from Type Ia supernovae, the cosmic microwave background temperature and polarization spectra, and baryon acoustic oscillations. We can use constraints from these datasets to predict what future probes of structure growth should observe. We apply this method to predict what range of cosmic shear power spectra would be expected if we lived in a  $\Lambda$CDM universe, with or without spatial curvature, and what results would be inconsistent and therefore falsify the model. Though predictions are relaxed if one allows for an arbitrary quintessence equation of state $-1\le w(z)\le 1$, we find that any observation that rules out $\Lambda$CDM due to excess lensing will also rule out all quintessence models, with or without early dark energy. We further explore how uncertainties in the nonlinear matter power spectrum, e.g.~from approximate fitting formulas such as Halofit, warm dark matter, or baryons, impact these limits.

\end{abstract}

\maketitle

\section{Introduction}

Consistency relations between cosmological observables exist for any underlying physical model class \cite{Mortonson:2008qy,Mortonson:2009hk}. This means that the combination of observables pertaining to the expansion history and those pertaining to structure growth could potentially falsify a given dark energy paradigm, such as the standard $\Lambda$CDM model of cold dark matter and a cosmological constant with Gaussian initial conditions, or smooth dark energy models with equation of state parameter $-1\le w(z)\le 1$, known as quintessence models.
For instance, even a single massive high-redshift cluster could falsify all \lcdm\ and quintessence models, if its mass falls significantly outside of what we predict based on Type Ia supernovae (SNe), the cosmic microwave background (CMB), baryon acoustic oscillations (BAO), and the local measurement of the Hubble constant ($H_0$) \cite{Mortonson:2010mj}.

Weak gravitational lensing, whereby distant galaxy images are distorted due to the gravitational effects of matter lying along the line of sight, is another key probe of cosmic structure growth, with promising results in recent years, e.g.~\cite{Fu:2007qq,Kilbinger:2008gk,Massey:2007gh,Massey:2007wb,Schrabback:2009ba}. 
See Refs.~\cite{Takada:2003ef,Hoekstra:2008db,Massey:2010hh} for recent reviews. However, current cosmological constraints are very limited; since we do not know the intrinsic shapes of galaxies, weak lensing is by its very nature a statistical measure and thus the power of a survey is directly related to its sky coverage \cite{Hu:1998az,Amara:2006kp}. Given future large-area surveys on the horizon, for instance from the ground with the Dark Energy Survey (DES) \cite{des_url} and the Large Synoptic Survey Telescope (LSST) \cite{lsst_url} or from space with {\it Euclid} \cite{euclid_url} and the Wide-Field Infrared Survey Telescope ({\it WFIRST}) \cite{wfirst_url}, it is particularly timely to determine our expectations \cite{Weinberg:2012es}.

Using the wealth of data in hand from distance measures and the CMB, we expand upon Ref.~\cite{Mortonson:2010mj} to predict these weak lensing observables within the
$\Lambda$CDM and quintessence paradigms. We focus on the shear power spectrum and show how the predictions relax as we generalize the model beyond flat $\Lambda$CDM and allow for curvature, an arbitrary dark energy equation of state, and early dark energy (EDE). Once the upcoming large-area weak lensing surveys are completed, we can compare their results to these predictions and possibly falsify the $\Lambda$CDM model or perhaps the entire quintessence paradigm. In this way weak lensing could provide a smoking gun for new physics such as primordial non-Gaussianity, phantom dark energy ($w(z)<-1$), sub-horizon dark energy clustering,  an interacting dark sector, or even a modification to general relativity.

This paper is organized as follows. In \S \ref{sec:methadology} we describe our methodology, including the datasets we use, the parameter spaces we explore, our Markov Chain Monte Carlo (MCMC) likelihood analysis, and how we compute the ingredients needed to predict the weak lensing shear power spectrum and two point correlation functions (2PCFs) for source redshift distributions typical of current and future surveys. In \S \ref{sec:wl} we present our weak lensing predictions and discuss how they depend on source redshift and dark energy model class. In \S \ref{sec:uncertainties} we explore uncertainties related to the matter power spectrum and SN light-curve fitting. We discuss and conclude in \S \ref{sec:discussion}.

\section{Methodology}
\label{sec:methadology}

Our methodology is based on that of Refs.~\cite{Mortonson:2008qy,Mortonson:2009hk,Mortonson:2010mj}. Briefly, we take current CMB constraints on the initial power spectrum plus current data related to the overall geometry and expansion history of the Universe, determine parameter constraints within a class of dark energy models using an MCMC analysis, and then compute the weak lensing predictions which can be used to falsify the class. The observational data we use come from local distance measures of $H_0$, the Type Ia SNe distance-redshift relation, BAO distance measures, and the CMB temperature and polarization power spectra.

\subsection{Data sets}
\label{sec:data}

The Type Ia SN sample we use is the compilation of 288 SNe from
Ref.~\cite{SDSS_SN}, consisting of data from the first season of the Sloan
Digital Sky Survey-II (SDSS-II) Supernova Survey, the ESSENCE survey
\cite{WoodVasey_2007}, the Supernova Legacy Survey \cite{Astier}, Hubble Space
Telescope SN observations \cite{Riess_2006}, and a collection of nearby SN
data \cite{Jha:2006fm}.  The light curves of these SNe have been uniformly
analyzed by \cite{SDSS_SN} using both the MLCS2k2 \cite{Jha:2006fm} 
and SALT2 \cite{Guy:2007dv} methods.
We use the SALT2 method for our default analysis but discuss the impact of
switching to MLCS2k2 in \S \ref{sec:sn}.

For the CMB, we use the most recent, 7-year release of data from the {\it WMAP}
satellite (WMAP7) \cite{Larson:2010gs} employing a modified version of the likelihood
code available at the LAMBDA web site \cite{WMAP_like} which is substantially
faster than the standard version while remaining sufficiently 
accurate \cite{Dvorkin:2010dn,wmapfast_url}.  We
compute the CMB angular power spectra using the code CAMB
\cite{Lewis:1999bs,camb_url} modified with the parametrized post-Friedmann
(PPF) dark energy module \cite{PPF,ppf_url} to include models with general
dark energy equation of state evolution where $w(z)$ may cross $w=-1$.

We use the BAO constraints from Ref.~\cite{PercivalBAO}, which combines 
data from SDSS and the 2-degree Field Galaxy Redshift Survey that
determine the ratio of the sound horizon at last scattering to the 
quantity $D_V(z)\equiv [zD^2(z)/H(z)]^{1/3}$ at redshifts $z=0.2$ and $z=0.35$.
Since these constraints actually come from galaxies spread over a range 
of redshifts, and our most general dark energy model classes allow the 
possibility of high frequency variations in the expansion rate $H(z)$ 
across this
range, we implement the constraints by taking the volume average of $D_V$ 
over $0.1<z<0.26$ (for $z=0.2$) and $0.2<z<0.45$ (for $z=0.35$).

Finally, we include the recent Hubble constant measurement from the SHOES team
\cite{SHOES}, based on SN distances at $0.023<z<0.1$ that are linked to a
maser-determined absolute distance using Cepheids observed in both the maser
galaxy and nearby galaxies hosting Type Ia SNe.  The SHOES measurement
determines the absolute distance to a mean SN redshift of $z=0.04$
which we implement as 
$D(z=0.04) = 0.04c/(74.2 \pm 3.6$~km~s$^{-1}$~Mpc$^{-1}$).

\subsection{Parameter constraints}

For a given parameter set $\bm{\theta}$ that  defines the cosmological model class, we use a modified version of the CosmoMC code \cite{Lewis:2002ah,cosmomc_url} to sample from the joint posterior distribution,
\be
{\cal P}(\bm{\theta}|{\bf x})=\frac{{\cal L}({\bf x}|\bm{\theta}){\cal P}(\bm{\theta})}{\int d\bm{\theta}{\cal L}({\bf x}|\bm{\theta}){\cal P}(\bm{\theta})}~,
\label{eq:bayes}
\ee
where ${\cal L}({\bf x}|\bm{\theta})$ is the likelihood of the dataset ${\bf x}$ given the model parameters $\bm{\theta}$ and ${\cal P}(\bm{\theta})$ is the prior probability density. We use the same priors as in \cite{Mortonson:2010mj}, namely flat priors that are wide enough to not limit the MCMC constraints from the aforementioned datasets. The most restrictive parameter class we will consider is that of a flat $\Lambda$CDM model,
\be
\bm{\theta}_{0} = \{\Omega_{\rm{b}}h^2, \Omega_{\rm{c}}h^2, \tau, \theta_A, n_s, \ln A_s\}~,
\label{eq:lcdmpar}
\ee
where $\Omega_{\rm{b}}h^2$ and $\Omega_{\rm{c}}h^2$ are the present physical baryon and cold dark matter densities relative to the critical density, 
$\tau$ is the reionization optical depth, $\theta_A$ is the angular size of 
the acoustic scale at last scattering, $n_{s}$ is the spectral index of the power spectrum of initial fluctuations, and $A_s$ is the amplitude of the initial curvature power spectrum at $k_{\rm p}=0.05~{\rm Mpc}^{-1}$.
All other parameters, including the Hubble constant $H_0 = 100\,h~{\rm km/s/Mpc}$, the present total matter density $\Omega_{\rm m}$ and dark energy density $\Omega_{\rm{DE}}$, and the amplitude of the matter power spectrum today $\sigma_8$, can be derived from this basic set.

We generalize this basic model, and consequently expand the aforementioned parameter set, in several ways. The first is to allow the spatial curvature $\Omega_{\rm{K}}=1-\Omega_{\rm{tot}}$ to vary. We further allow for quintessence by decomposing the dark energy equation of state into principal components (PCs) for $z<1.7$,
\be
w(z) + 1  = \sum_{i=1}^{N_{\rm{max}}} \alpha_i e_i(z)~,
\label{eq:pcstow}
\ee
where the equation of state at higher redshift is a constant $w(z > 1.7) = w_{\infty}$, the $\alpha_i$ are the PC amplitudes, the $e_i(z)$ are the PC functions, and $N_{\rm{max}}=10$ has been found to provide a complete basis for our purposes here \cite{Mortonson:2009hk}. 
The PCs we use here are the eigenfunctions of the Fisher matrix for the 
combination of CMB constraints from {\it Planck} and {\it WFIRST}-like 
SN data, as described in Ref.~\cite{Mortonson:2008qy}.
We will fix $w_{\infty}=-1$ for the model classes which exclude EDE, as a constant equation of state dark energy component quickly becomes negligible at early times, and allow it to vary otherwise with a flat prior on $e^{w_{\infty}}$ between $e^{-1}$ and 1. A detailed discussion of this EDE approach and its relation to CMB
constraints can be found in Appendix B of Ref.~\cite{Mortonson:2010mj}.

Quintessence models describe dark energy as a scalar field with kinetic and potential contributions to energy and pressure.  Barring models where large kinetic and (negative)
potential contributions cancel (e.g.~\cite{Mortonson:2009qq}), quintessence equations of state are restricted to $-1\leq w(z)\leq 1$. Following 
Ref.~\cite{Mortonson:2008qy}, this bound is conservatively implemented
with uncorrelated top-hat priors on the PC amplitudes $\alpha_i$.  
Any combination of PC amplitudes that is
rejected by these priors must arise from an equation of state $w(z)$ that
violates the bound on $w(z)$, but not all models that are allowed by the priors
strictly satisfy this bound.

In summary, the full set of parameters we will consider here are 
\be
\bm{\theta} = \{\bm{\theta}_{0} , \Omega_{\rm{K}}, \alpha_1,\ldots, \alpha_{10}, e^{w_{\infty}}\}~,
\label{eq:parametersfull}
\ee
where we will look at models of increasing generality by exploring increasingly larger subsets of this parameter set. 

\subsection{Observables}

Once we have sampled the joint posterior distribution of the cosmological parameters, we can then compute the posterior probabilities for any derived statistic, in particular cosmic shear observables. 
As an intermediate step, it is useful to first consider the two basic ingredients that the
two-point shear observables are constructed from: the comoving angular diameter distance
$D$ and the nonlinear matter power spectrum $\Delta^2_{\rm NL}$.  
The former is related to the cosmological parameters through 
\be
D(\chi)=\frac{1}{\sqrt{|\Omega_{\rm{K}}|H_0^2}}S_{\rm{K}}\left[\chi\sqrt{|\Omega_{\rm{K}}|H_0^2}\right]~,
\ee
where $S_{\rm{K}}(x)$ equals $x$ in a flat universe ($\Omega_{\rm{K}}=0$), $\sinh x$ in an open universe ($\Omega_{\rm{K}}>0$), and $\sin x$ in a closed universe ($\Omega_{\rm{K}}<0$).  Here the comoving radial coordinate is
\be
\chi(z) = \int_0^z\frac{dz'}{H(z')},
\ee
where the Hubble expansion rate is
\be
H(z)=H_0\left[\Omega_{\rm{m}}(1+z)^3+\Omega_{\rm{DE}}f(z)+\Omega_{\rm{K}}(1+z)^2\right]^{1/2}
\ee
with
\be
f(z)=\exp\left[3\int^z_0 dz'\frac{1+w(z')}{1+z'}\right]\,,
\label{eqn:fz}
\ee
and the contribution from radiation is assumed to be negligible.

The linear matter power spectrum depends on the initial matter power spectrum 
and the growth function of linear density perturbations
$\delta\propto G a$, where $a$ is the scale factor.  Given the smooth dark energy paradigm, the growth function is given by
\be
G'' + \left(4+\frac{H'}{H}\right)G'+\left[3+\frac{H'}{H}-\frac{3\Omega_{\rm{m}} H_0^2(1+z)^3}{2 H^2(z)}\right]G = 0~,
\ee
where primes denote derivates with respect to $\ln a$ and the function is normalized so
that $G=1$ in the matter dominated limit.   
We then compute the linear power spectrum at redshift $z$ by rescaling the 
$z=0$ spectrum $\Delta_{\rm L}^2(k;0)$ (computed using CAMB with the 
PPF dark energy module) by the growth function:
\begin{equation}
\Delta_{\rm L}^2(k; z) = \left[ {G(z) \over (1+z)G(0) } \right]^2 \Delta_{\rm L}^2(k; 0) \,.
\end{equation}

We shall see that current CMB constraints on  $\Omega_{\rm{m}} h^2$ still allow substantial variation in the matter radiation equality scale and hence the
shape of the linear power spectrum.

Cosmic shear observables however depend on the full nonlinear power spectrum.
We  compute this quantity using the Halofit fitting function \cite{Smith:2002dz}, which we will now describe. 
The halo model decomposes the nonlinear power spectrum into the sum of two contributions,
\be
\Delta_{\rm{NL}}^2(k) = \Delta_{\rm{Q}}^2(k) + \Delta_{\rm{H}}^2(k)~,
\ee
where $\Delta_{\rm{Q}}^2(k)$ is the ``two-halo" term, which encapsulates quasi-linear power from large-scale halo placement, and $\Delta_{\rm{H}}^2(k)$ is the ``one-halo" term, which arises due to correlations within the haloes themselves. The Halofit fitting functions depend on parameters based on Gaussian filtering of the linear power spectrum with variance
\be
\sigma^2(R)\equiv \int \Delta_{\rm{L}}^2(k)\exp\left(-k^2R^2\right)d\ln k~:
\ee
the nonlinear scale $k_{\rm{NL}}$ defined such that $\sigma\left(k_{\rm{NL}}^{-1}\right) \equiv 1$, the effective spectral index,
\be
n \equiv -3 -\left.\frac{d\ln\sigma^2(R)}{d\ln R}\right|_{\sigma=1}~,
\ee
and the spectral curvature,
\be
C\equiv -\left.\frac{d^2\ln\sigma^2(R)}{d\ln R^2}\right|_{\sigma=1}~.
\ee
Then the nonlinear power spectrum is parameterized by a set of coefficients $(a_n,b_n,c_n,\gamma_n,\alpha_n,\beta_n,\mu_n,\nu_n)$ which are allowed to vary as a function of the aforementioned spectral index and curvature so as to fit N-body simulation data. In terms of these coefficients, 
\be 
\Delta^2_{\rm{Q}}(k)= \Delta^2_{\rm{L}}(k) \left\{\frac{\left[1+\Delta_{\rm{L}}^2(k)\right]^{\beta_n}}{1+\alpha_n\Delta^2_{\rm{L}}(k)} \right\}\exp{[-f(y)]}~,
\ee
where $y\equiv k/k_{\rm{NL}}$ and $f(y)=y/4+y^2/8$ is a function introduced to truncate at high $k$, and 
\be 
\Delta^2_{\rm{H}}(k) = \frac{\Delta^{2\ \prime}_{\rm{H}}(k)}{1+\mu_ny^{-1}+\nu_ny^{-2}}~,
\ee
with
\be 
\Delta^{2\ \prime}_{\rm{H}}(k) = \frac{a_n y^{3f_1(\Omega_{\rm{m}})}}{1+b_ny^{f_2(\Omega_{\rm{m}})}+\left[ c_nf_3(\Omega_{\rm{m}}) y\right]^{3-\gamma_n}}~.
\label{deltah}
\ee
The coefficients are
\ba
\log_{10} a_n &=& 1.4861 + 1.8369n + 1.6762n^2 + 0.7940n^3\nonumber\\ 
& &+ 0.1670n^4 -0.6206C~,\nonumber\\
\log_{10} b_n &=& 0.9463 + 0.9466n + 0.3084n^2 - 0.9400C~,\nonumber\\
\log_{10} c_n &=& -0.2807 + 0.6669n + 0.3214n^2 - 0.0793C~,\nonumber\\
\gamma_n &=& 0.8649 + 0.2989n + 0.1631C~,\nonumber\\
\alpha_n &=& 1.3884 + 0.3700n - 0.1452n^2~,\nonumber\\
\beta_n &=& 0.8291 + 0.9854n + 0.3401n^2~,\nonumber\\
\log_{10}\mu_n &=& -3.5442+0.1908n~,\nonumber\\
\log_{10}\nu_n &=& 0.9589+1.2857n~.
\ea
The functions $(f_1,f_2,f_3)$ are power laws, with the exponents fit to N-body data for either matter-only open models
\be 
\left. 
\begin{array}{lll}
f_{1}(\Omega_{\rm m}) & = & \Omega_{\rm m}^{\;-0.0732} \\
f_{2}(\Omega_{\rm m}) & = & \Omega_{\rm m}^{\;-0.1423}\\
f_{3}(\Omega_{\rm m}) & = & \Omega_{\rm m}^{\;0.0725}\\
\end{array}
\right\} \ \ \Omega_{\rm m}\le1
\ee
or flat $\Lambda$CDM models
\be 
\left. 
\begin{array}{lll}
f_{1}(\Omega_{\rm m}) & = & \Omega_{\rm m}^{\;-0.0307}\\
f_{2}(\Omega_{\rm m}) & = & \Omega_{\rm m}^{\;-0.0585}\\
f_{3}(\Omega_{\rm m}) & = & \Omega_{\rm m}^{\;0.0743}\\
\end{array}
\right\} \ \ \Omega_{\rm m}+\Omega_{\rm DE}=1~.
\ee
Note that we will use these fitting functions for all of our model classes, despite that fact that they were calibrated on simulations where the dark energy equation of state never deviated from $w=-1$.
We comment on this simplification in \S \ref{sec:nonlinearpk}.  For our main result that
\lcdm\ sets an upper bound on shear statistics, we expect that this approximation suffices.

As suggested in Appendix C of Ref.~\cite{Smith:2002dz}, we use linear interpolation for models in which $\Omega_{\rm{DE}}$ is neither zero nor $1-\Omega_{\rm{m}}$. We further use the high-$k$ correction \cite{halofit_correction_url} 
\be
\Delta^2_{\rm NL}(k)-\Delta^2_{\rm{L}}(k) \rightarrow \left[\Delta^2_{\rm NL}(k)-\Delta^2_{\rm{L}}(k)\right]\left(\frac{1+2x^2}{1+x^2}\right)~,
\label{halofitcorrection}
\ee
where $x\equiv k/(10\,h~{\rm Mpc}^{-1})$.
These fitting functions have been found to be inaccurate (even for flat $\Lambda$CDM) at up to the $5$--$10\%$ level, for instance with the Coyote universe project \cite{Heitmann:2008eq,Heitmann:2009cu,Lawrence:2009uk}. In \S \ref{sec:uncertainties} we will explore how our predictions depend on the accuracy of the one-halo term amplitude $a_n$. We will also show (and exploit) how one can use $c_n$ to parameterize uncertainties due to warm dark matter and baryonic processes.

From the distance-redshift relation and the nonlinear matter power spectrum, we can then compute the shear power spectrum, equal to the convergence power spectrum
\be
{l^2 P_{\kappa} \over 2\pi} = {9\pi \over 4 c^4 l}  {\Omega_{\rm{m}}^2H_0^4  } \int_0^{\infty}dz {D^3 \over H} \frac{g^2(z)}{a^2}\Delta^2_{\rm NL}\left(\frac{l}{D(\chi)},z\right)~,
\label{shearpower}
\ee
where $k \approx l/D$ in units of Mpc$^{-1}$ in the Limber approximation and we have defined the geometric lensing efficiency factor
\be
g(z)\equiv\int^{\infty}_{z} dz' n(z')\frac{D(\chi'-\chi)}{D(\chi')}~.
\ee
The efficiency factor is weighted according to the source distribution in a given survey, $n(z)$, normalized such that $\int_0^{\infty}n(z) dz =1$. For this paper we will use the simple model
\be
n(z)\propto  \left(\frac{z}{z_0}\right)^{\alpha}\exp\left[-\left(\frac{z}{z_0}\right)^{\beta}\right]~,
\ee
with parameters $z_0$, $\alpha$, and $\beta$.  This parameterization is similar to what has been used in both the COSMOS \cite{Massey:2007gh} and CFHTLS \cite{Benjamin:2007ys} analyses, with $(z_0,\alpha,\beta)=(0.894,2.0,1.5)$ for COSMOS and $(z_0,\alpha,\beta)=(0.555,1.197,1.193)$ for CFHTLS. This leads to approximate median redshifts of $1.3$ and $0.8$, respectively, for our simplified COSMOS and CFHTLS surveys.
Note that these distributions closely approximate the ones
expected for {\it WFIRST} \cite{Jouvel:2009mh,Green:2011zi} and DES. 
In the work that follows we will plot results for both of these simple source distributions, but will specialize to CFHTLS (as in \cite{Eifler:2010kt}) when results for the two are 
similar.

Finally, we also compute the 2PCFs 
 \be
\xi_{+/-}(\theta) = \frac{1}{2\pi}\int_0^{\infty}dl \, l \, J_{0/4}(l\theta)P_{\kappa}(l)~
\ee
of the shear components from the power spectrum.  Being the real-space correlation
and the easier quantity to measure on small scales, the 2PCFs are better suited to compare against current measurements than $P_\kappa$ itself.

\section{Weak lensing predictions}
\label{sec:wl}

Current distance constraints are highly predictive for growth of structure statistics such as
the weak lensing power spectrum in the flat $\Lambda$CDM context.
Since this is the standard model of cosmology and our baseline case, we shall use it to illustrate the steps in our chain of inference in \S \ref{sec:lcdm}.  We then discuss observations that would falsify 
$\Lambda$CDM in favor of quintessence and those that would falsify the whole quintessence class
of $-1 \le w(z)  \le 1$ smooth dark energy models in \S \ref{sec:quintessence}.

\subsection{$\Lambda$CDM}
\label{sec:lcdm}

\begin{figure}[t]
\includegraphics[width=8cm]{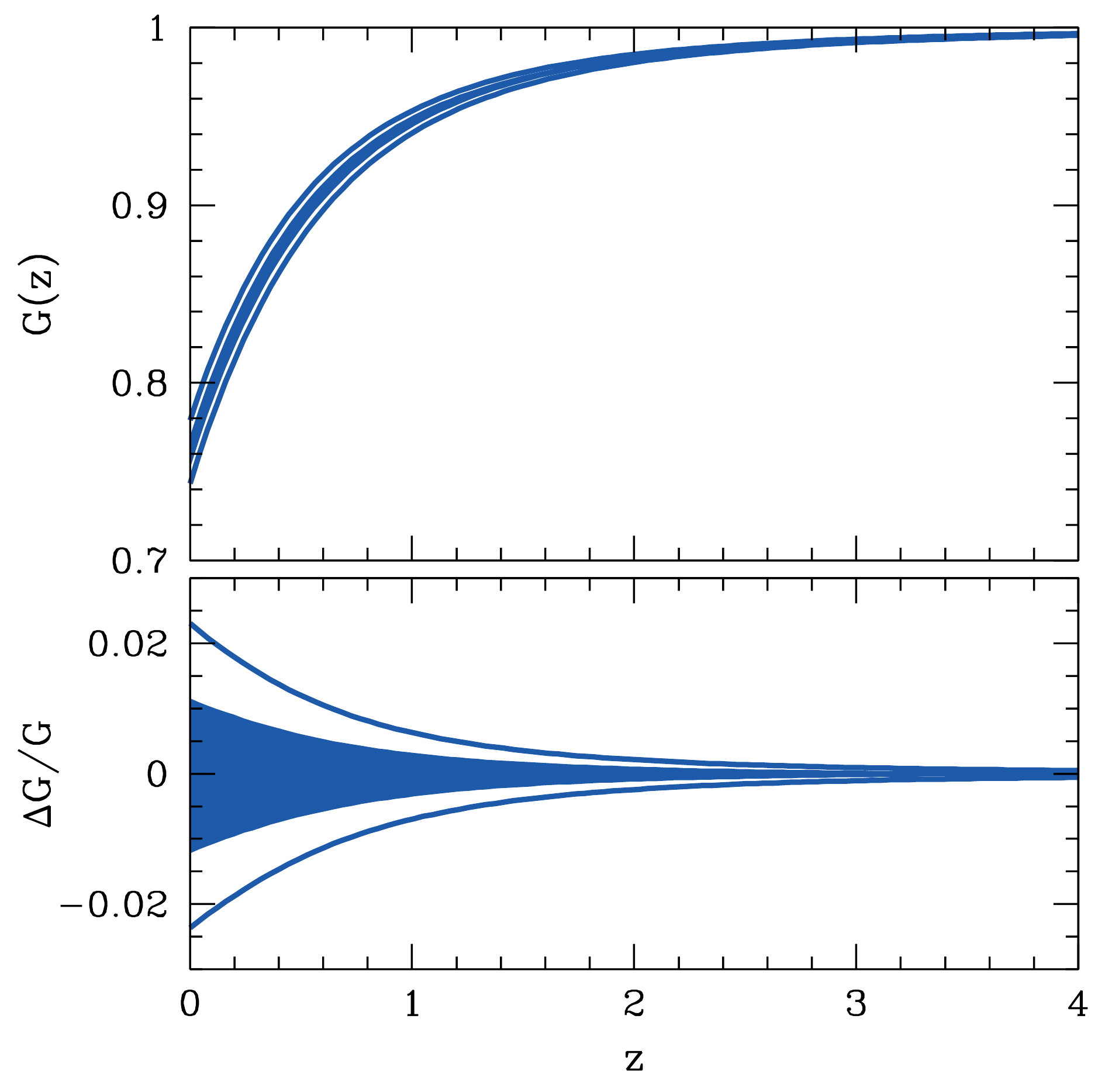}
\caption{Flat $\Lambda$CDM prediction for the growth factor (upper panel) and its deviation from the ML flat $\Lambda$CDM model (lower panel). 
The shaded regions correspond to 68\% CL and the curves bound the 95\% CL regions.}
\label{fig:growth}
\end{figure}

In Fig.~\ref{fig:growth}, we start with the growth function prediction.  
Unless stated otherwise, henceforth shaded regions will correspond to 68\% confidence level (CL) regions and curves will bound 95\% CL regions defined to have equal posterior probabilities in the two tails. 
We will often plot results as fractional differences from the prediction for the maximum likelihood (ML) model.
The growth predictions are most precise at high redshifts, but even at $z=0$ the allowed range in the growth function is only $1$--$2\%$ \cite{Mortonson:2009hk}.

\begin{figure}[t]
\includegraphics[width=8cm]{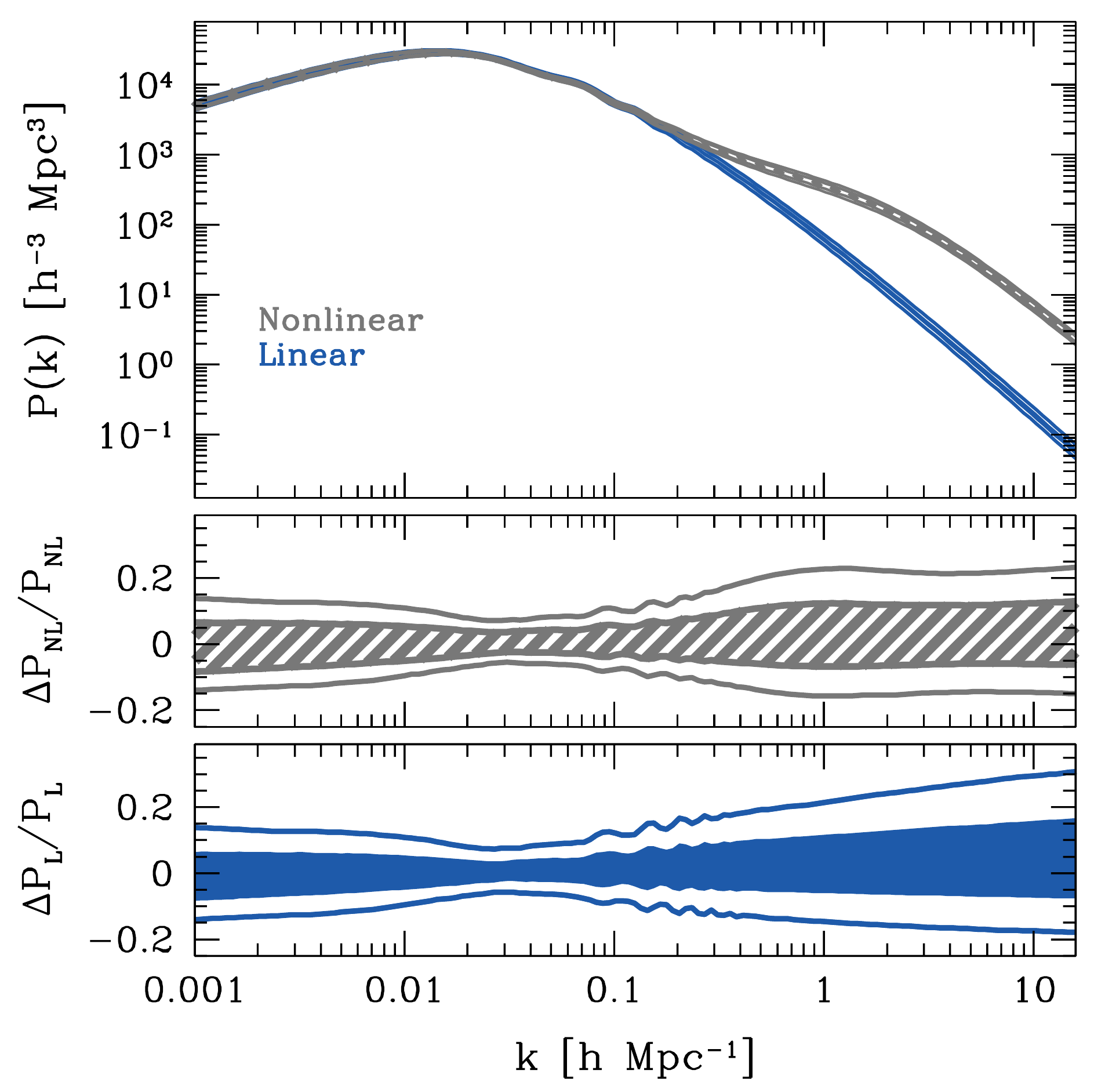}
\caption{Flat $\Lambda$CDM prediction for the $z=0$ matter power spectrum, showing the 68\% and 95\% CL regions as in Fig.~\ref{fig:growth}.
Upper panel: Linear (blue solid) and nonlinear (grey hatched) matter power spectra.
Middle panel: Fractional confidence range in the nonlinear spectrum around the  ML flat $\Lambda$CDM model.
Lower panel: The same for the linear power spectrum.   Note $k$ is in units of $h$ Mpc$^{-1}$.}
\label{fig:pklin}
\end{figure}

As described in the previous section, we can combine the growth function with the initial
power spectrum and transfer function to predict the linear matter power spectrum 
at $z=0$; see the upper and lower panels of Fig.~\ref{fig:pklin}.  
For plotting purposes we follow the usual convention of taking $P(k) = (2\pi^2/k^3)\Delta^2(k)$
with $k$ defined in $h$ Mpc$^{-1}$ for both linear and nonlinear power spectra.
Here the predictions
carry $\sim 10\%$ errors in spite of the precise growth results. 
The dominant
source of uncertainty is from the measurements of the 
matter density and shape of the initial power spectrum from WMAP7.  In
particular, uncertainties in $\Omega_{\rm{m}} h^2$ correspond to shifts in
matter-radiation equality which cause  left-right
shifts in the power spectrum.   There are also contributions from tilt ($n_s$) uncertainties
that pivot the spectrum
around the best constrained {\it WMAP} scale of $k \sim 0.02$ Mpc$^{-1} \approx 0.03 h$ Mpc$^{-1}$.    Both of these types of uncertainties should be reduced by a factor of a few with
{\it Planck} CMB data and allow the full precision of the growth predictions to be utilized
(see \S \ref{sec:linearpk} and Fig.~\ref{fig:planck}). 

Also shown in Fig.~\ref{fig:pklin} is the full nonlinear matter power spectrum as computed with Halofit 
(grey hatched curve in the upper panel), along with its deviation from the ML prediction in the middle panel.
For the nonlinear power spectrum, the uncertainties are the same as the linear one for 
$k < k_{\rm NL} \sim 0.3~h~\rm{Mpc}^{-1}$.   For smaller scales, the nonlinear power spectrum
uncertainties no longer reflect the linear uncertainties.    Whereas the tilt $n_s$ makes the
linear uncertainties continue to grow larger, the nonlinear ones become saturated 
reflecting the fixed nature of the one-halo piece of the Halofit prescription. We shall see that uncertainties here
are dominated by the accuracy of Halofit and the ability of adjustments in its parameters to model baryonic physics (see \S \ref{sec:nonlinearpk}).  

\begin{figure}[t]
\centering
\includegraphics[width=8cm]{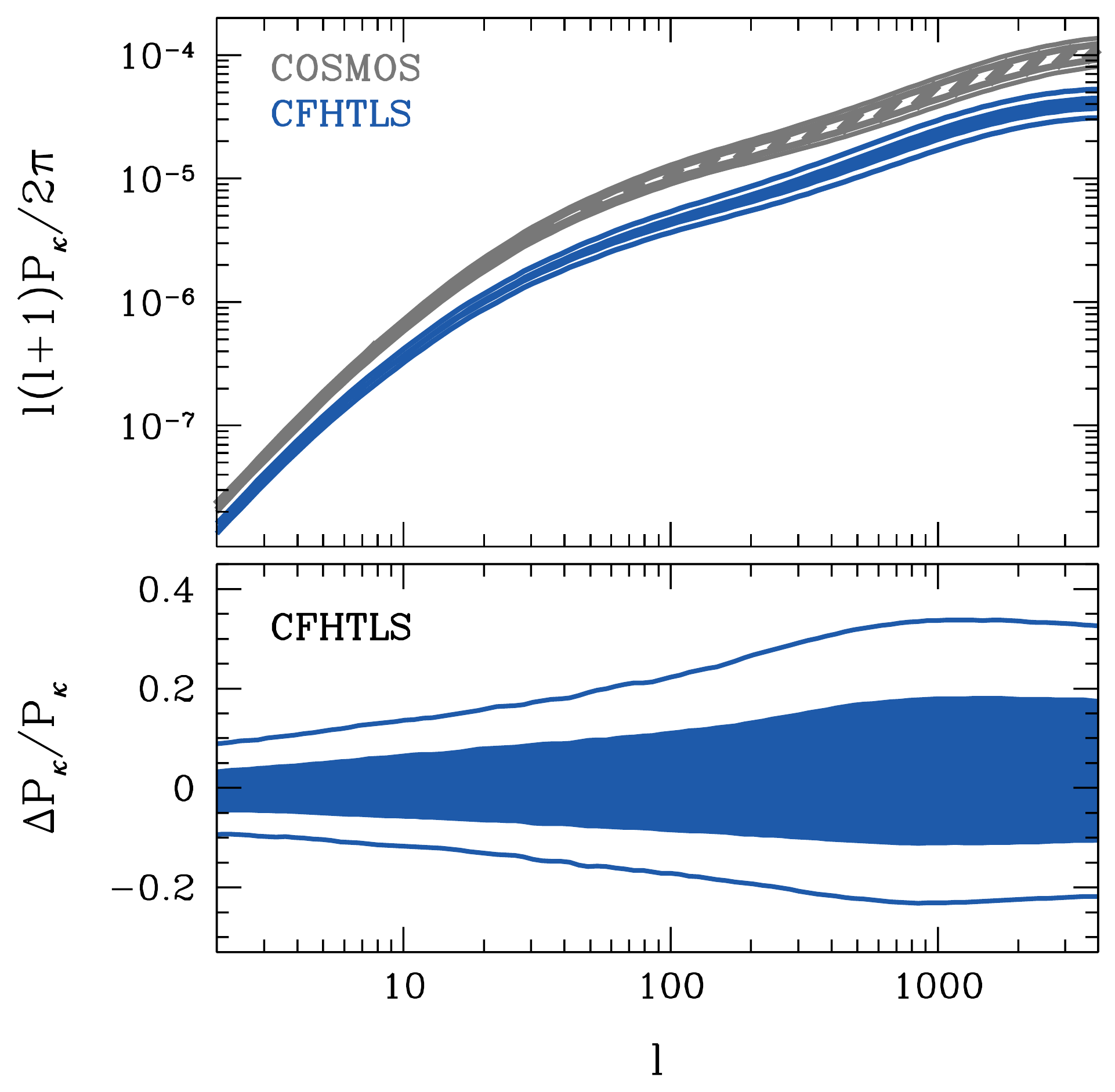}
\caption{Upper panel: Flat $\Lambda$CDM predictions for the shear power spectrum, showing the 68\% and 95\% CL regions as in Fig.~\ref{fig:growth} for COSMOS (upper, grey hatched) and CFHTLS (lower, blue solid). Lower panel: CFHTLS shear power spectrum prediction plotted with respect to the ML flat $\Lambda$CDM model.}
\label{fig:shear}
\end{figure}

Uncertainties in $P_{\rm NL}(k,z)$ propagate into the shear statistics.
We show the shear power spectrum as computed from Eq.~(\ref{shearpower})
in Fig.~\ref{fig:shear} for both the COSMOS and CFHTLS source distributions.
The confidence intervals mainly reflect those of the nonlinear matter power spectrum.   The interval is actually slightly narrower at low $l$ than at low $k$ in 
Fig.~\ref{fig:pklin}.  
Significant contributions to $P_\kappa$ come from $z>0.5$, where both linear growth function uncertainties diminish and 
the linear power spectrum of the gravitational potential is better fixed by the CMB (in the relevant units of Mpc$^{-1}$ rather than
in $h$ Mpc$^{-1}$).
Likewise, at a fixed angular scale CFHTLS predictions tend to be slightly weaker than those from COSMOS due to the lower source redshifts typical of ground-based surveys.

\begin{figure}[t]
\includegraphics[width=8cm]{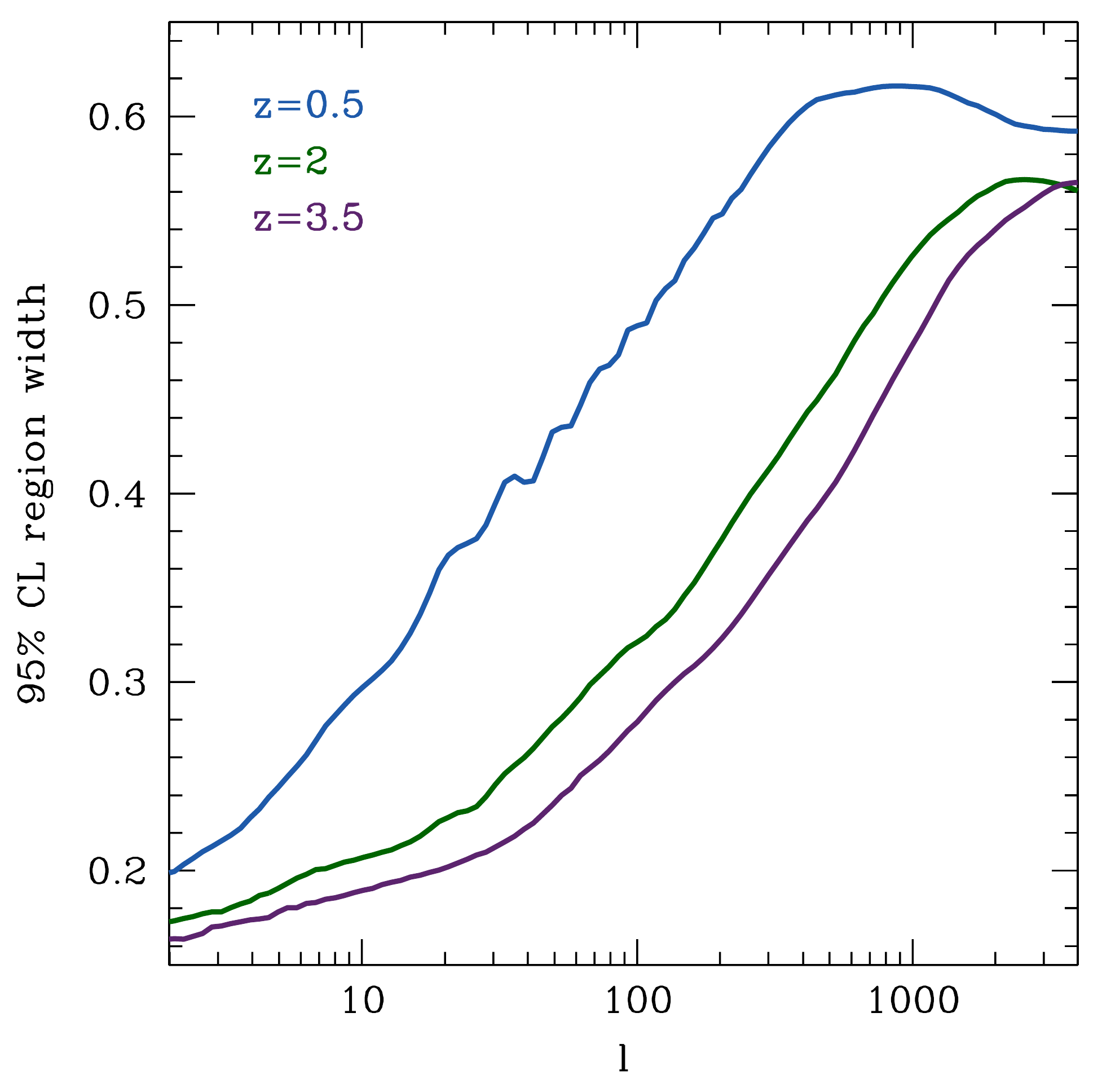}
\caption{Single source plane, 95\% CL full-width extent for the shear power spectrum $\Delta P_\kappa/P_\kappa$ (as plotted in the lower panel of Fig.~\ref{fig:shear}) as a function of $l$, for sources at $z=0.5$ (top), 2 (middle), and 3.5 (bottom).}
\label{fig:shear_z}
\end{figure}

We can make this redshift dependence explicit by replacing the realistic redshift distributions with
idealized single source planes.
Figure~\ref{fig:shear_z} shows the 95\% CL region widths for sources at $z=0.5$, 2, and 3.5.
Note that the well-constrained pivot in the power spectrum projects to higher multipole
at higher redshift leading to tighter predictions at a fixed multipole. 
Given the tighter predictions, high redshift cosmic shear measurements provide an interesting opportunity to falsify the flat $\Lambda$CDM model.

\begin{figure}[t]
\includegraphics[width=8cm]{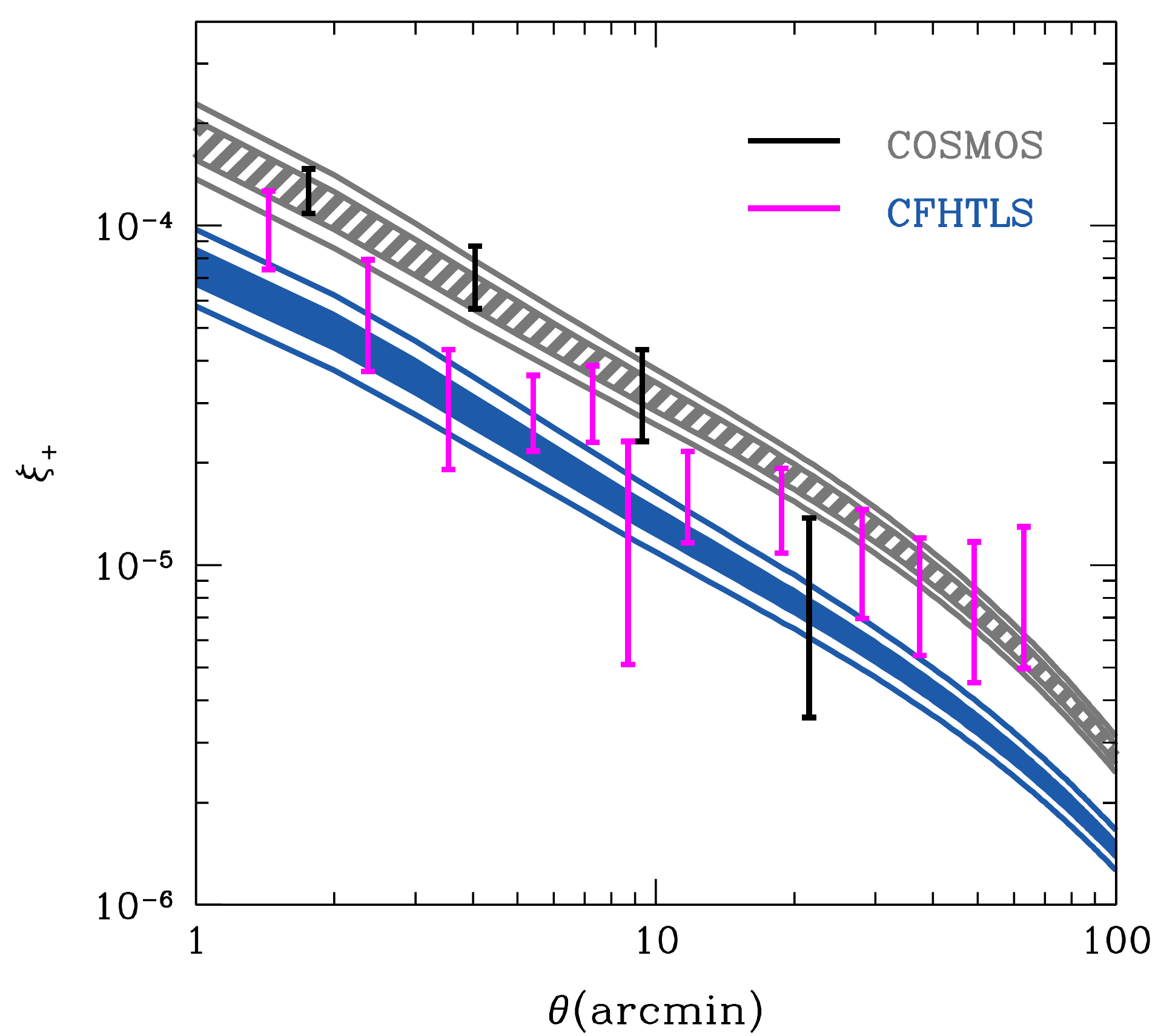}
\caption{Flat $\Lambda$CDM predictions for $\xi_{+}$, showing the 68\% and 95\% CL regions as in Fig.~\ref{fig:shear} for both the COSMOS (grey hatched) and CFHTLS (blue solid) source redshift distributions. Also shown are the data points from Refs.~\cite{Schrabback:2009ba} (black points) and \cite{Benjamin:2007ys} (magenta points) with $1\sigma$ error bars. Note that the error bars at different angular separations are correlated.}
\label{fig:xip_comp}
\end{figure}

In Fig.~\ref{fig:xip_comp} we show the 2PCF $\xi_+$ which is more useful for comparison with the relevant observations from COSMOS \cite{Schrabback:2009ba} and CFHTLS \cite{Benjamin:2007ys}. The displayed $1\sigma$ error bars are computed from the full covariance matrices estimated for each survey, as described in Refs.~\cite{Schrabback:2009ba,Benjamin:2007ys}. The predicted range of flat $\Lambda$CDM models appears to be consistent with the observations. However, the error bars at different angular scales are heavily correlated, and therefore do not represent the actual uncertainty at any individual scale. Further note that these small-volume surveys are not well-suited for making statements for or against ruling out the $\Lambda$CDM model; COSMOS results
use a 1.64 ${\rm deg}^2$ field containing 76 galaxies per ${\rm arcmin}^2$, and CFHTLS results use 22 ${\rm deg}^2$ containing 12 galaxies per ${\rm arcmin}^2$.

If future observations falsify these predictions, then one would need to generalize the
cosmological model class.
The next simplest class of models retains $\Lambda$ as the dark energy but allows
for non-vanishing spatial curvature  $\Omega_{\rm{K}}$ in the $\Lambda$CDM context.  We find a minimal error increase in this class, as shown in the upper panel of Fig.~\ref{fig:shear_gen}.  This is because spatial curvature is well constrained in the $\Lambda$CDM paradigm. Thus a measurement that falsifies the flat $\Lambda$CDM model would also falsify the $\Lambda$CDM assumption itself, indicating that the dark sector is more complicated.

\subsection{Quintessence}
\label{sec:quintessence}

Measurements of shear observables outside the bounds shown in the previous subsection would be in statistical conflict with $\Lambda$CDM. Barring systematic errors and unknown astrophysical effects, some of which we address in \S \ref{sec:uncertainties},
such a measurement would indicate that the true cosmology belongs to a wider class of models.  

\begin{figure}[t]
\includegraphics[width=8cm]{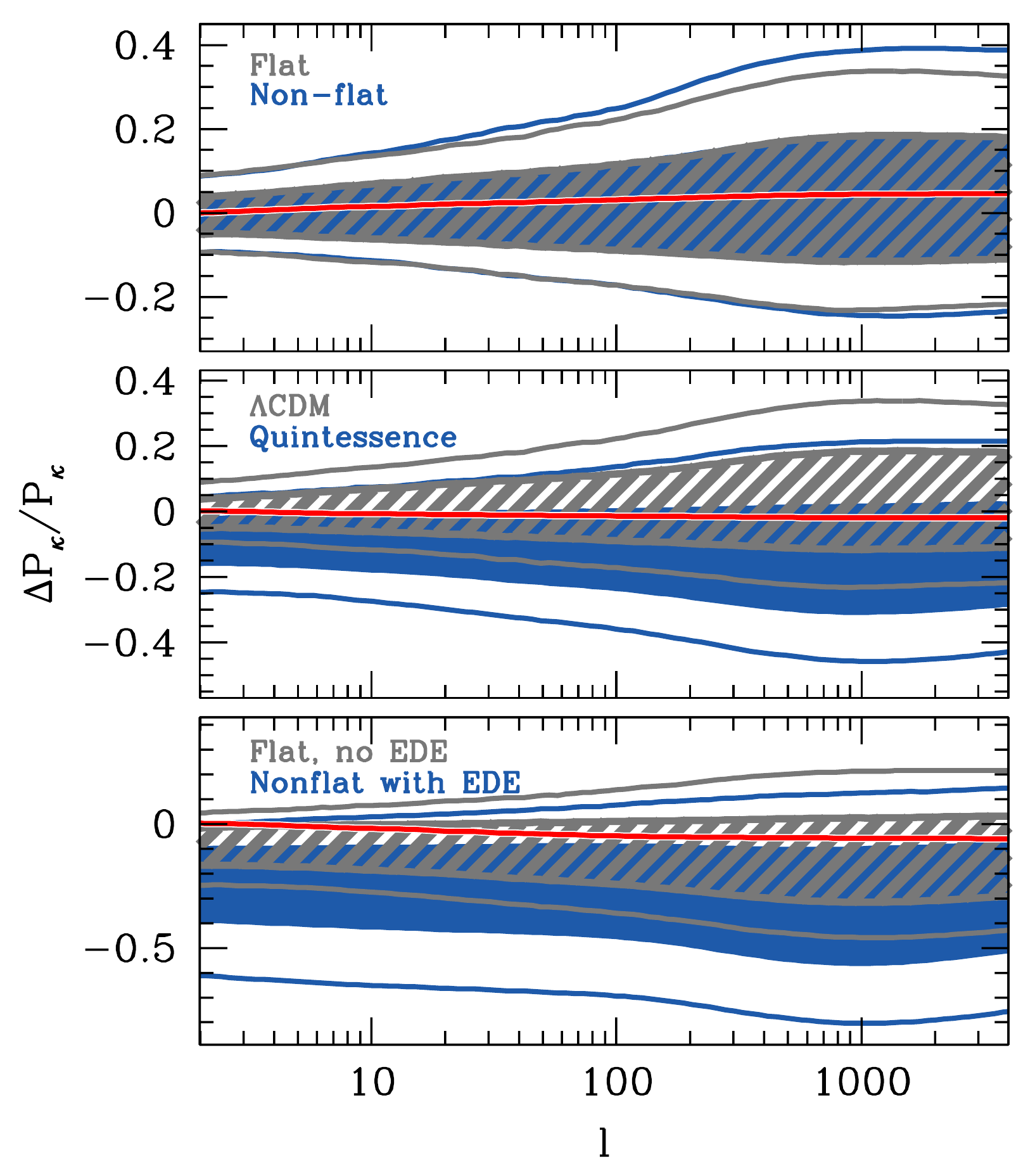}
\caption{Generalizing the model class for shear power spectrum predictions for the CFHTLS source redshift distribution.   In all cases, the range is shown relative to 
the ML flat $\Lambda$CDM model with blue regions indicating the wider model class
and gray hatched regions the narrower: (upper)
 $\Lambda$CDM flat vs non-flat predictions; (middle) flat $\Lambda$CDM vs  flat quintessence without EDE; (lower) flat quintessence without EDE vs  nonflat quintessence with EDE (blue solid).
Red lines correspond to the maximum likelihood model of the more general of the two
classes compared.}
\label{fig:shear_gen}
\end{figure}


If we relax the equation of state to allow all 10 PC amplitudes to vary but revert to the flatness assumption, we find that the contours shift toward lower power as shown in the middle panel of Fig.~\ref{fig:shear_gen}.  
Nevertheless, the ML quintessence model is 
quite similar to the flat \lcdm\  ML model and is near the 68\% CL upper limit 
of the quintessence model class.    This is an 
artifact of the parameter priors:  given the freedom in $w(z)$ there are many ways to reduce
the growth in ways unconstrained by distance measures  (see Fig.~3 of \cite{Mortonson:2009hk}). The converse is not true as there are no models with good likelihood values
in the quintessence class with more power than already allowed in flat \lcdm, an effect also seen in Ref.~\cite{Mortonson:2010mj}.  The effective data-driven upper bound on shear statistics for this quintessence class remains that of flat \lcdm\ in spite of the downward shift in posterior contours.

The lower panel of Fig.~\ref{fig:shear_gen} overlays the results for flat quintessence with $w_{\infty}=-1$, and quintessence with both curvature and $w_{\infty}$ allowed to vary (the latter generalization potentially allowing for EDE). This additional freedom further reduces growth, in accordance with the results of Ref.~\cite{Mortonson:2010mj}. This is one of our main results, namely that generalizing from \lcdm\ to quintessence models only serves to reduce cosmic shear. 
We will show below in \S~\ref{sec:nonlinearpk} that warm dark matter could also only explain a power deficit; likewise, massive neutrinos would only decrease power.\footnote{Massive neutrinos would also shift parameter values due to their effect on the CMB, so accurately quantifying the suppression of the predicted shear power requires including neutrino mass as a parameter in the initial MCMC analysis, which we leave for future work.} It is possible for baryonic effects to increase power in either \lcdm\ or quintessence, but only at high multipoles.
In general, observations which rule out the $\Lambda$CDM model by finding a shear excess that cannot be explained by astrophysical uncertainties also falsify 
the entire
quintessence paradigm. 

\section{Systematic and Other Uncertainties}
\label{sec:uncertainties}

We now turn to exploring various factors that can impact the statistical predictions
shown in the previous section. These uncertainties largely stem from uncertainties in the matter power spectrum and systematic uncertainties in the SNe data. In this section we will discuss the impact of an improved initial linear matter power spectrum from {\it Planck} (\S \ref{sec:linearpk}), uncertainties in the nonlinear matter power spectrum from small-scale physics and the Halofit fitting function (\S \ref{sec:nonlinearpk}), tomography
(\S \ref{sec:tomography}), and supernova light-curve fitting (\S \ref{sec:sn}).

\subsection{Linear Matter Power Spectrum}
\label{sec:linearpk}

Much of the residual statistical uncertainty in the flat $\Lambda$CDM predictions comes from 
uncertainties in the linear matter power spectrum from WMAP7.   The main sources
are the matter density $\Omega_{\rm{m}} h^2$, tilt $n_s$ and normalization $\ln A_s$.    
These uncertainties should soon be improved by {\it Planck}. We show how the predictions change from importance sampling to {\it Planck}-like priors on the parameters $\Omega_{\rm{b}}h^2$, $\Omega_{\rm{m}}h^2$, $n_s$, and $\ln A_s$, using the DETF projected covariance matrix \cite{DETF}
and assuming the mean corresponds to the ML  flat $\Lambda$CDM model. 

\begin{figure}[t]
\centering
\includegraphics[width=8cm]{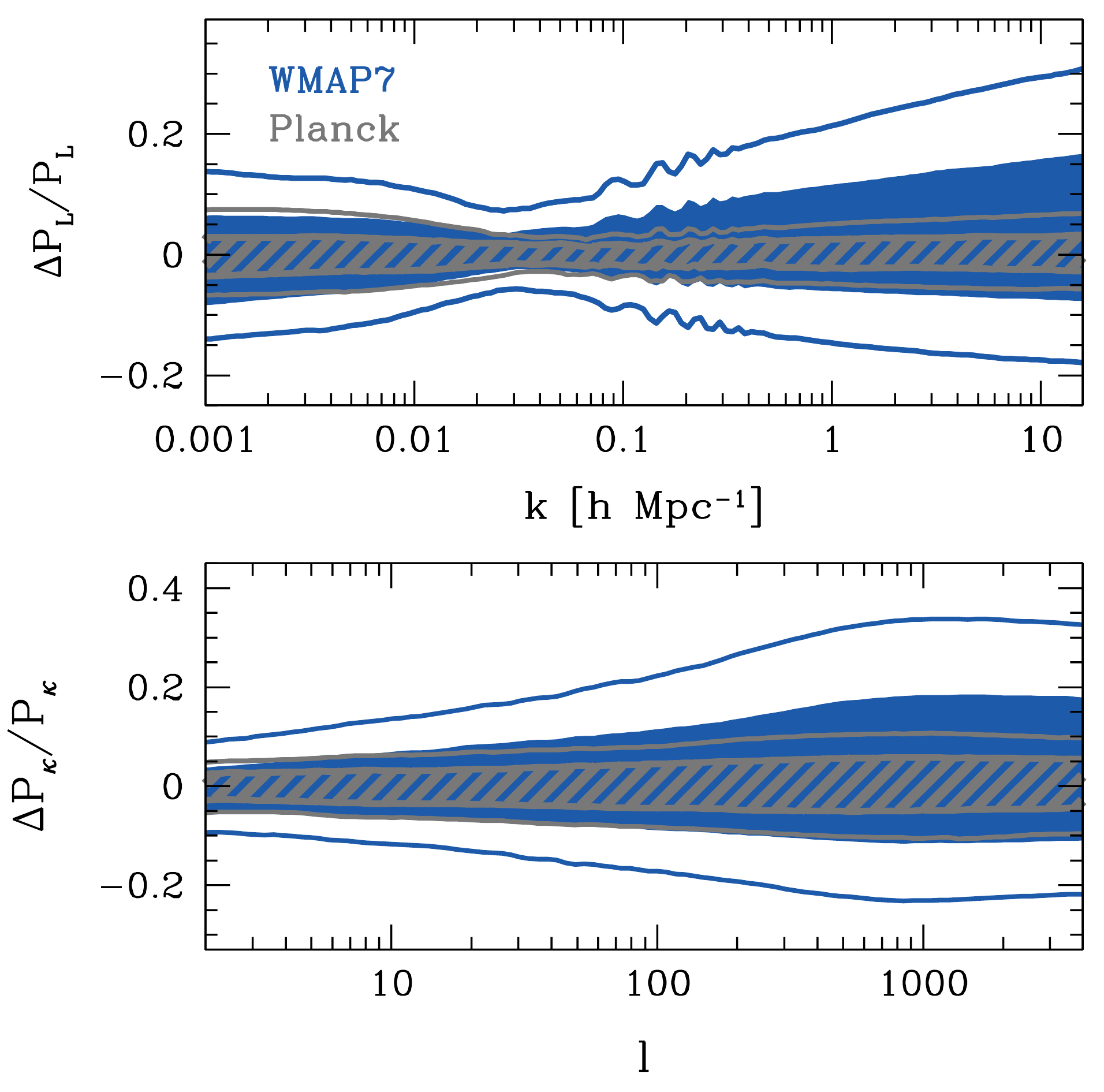}
\caption{Expected improvements from {\it Planck} (grey hatched) over WMAP7 (blue solid) 
for flat $\Lambda$CDM predictions.  Upper panel shows the linear matter power spectrum and the lower panel shows the shear power spectrum  for the CFHTLS source redshift distribution, with  the 68\% and 95\% CL regions relative to the ML flat \lcdm\ model.}
\label{fig:planck}
\end{figure}

In Fig.~\ref{fig:planck} we show the improvement in the fractional constraints (grey hatched curves) on the linear power spectrum (upper panel) and on the shear power spectrum of the CFHTLS source redshift distribution (lower panel).   One typically should expect Planck to improve the precision of predictions by more than a factor of 2 to the level of $\pm 3$--$5\%$, which will greatly enhance the prospects for 
falsifying the standard $\Lambda$CDM paradigm.

\begin{figure}[t]
\centering
\includegraphics[width=8cm]{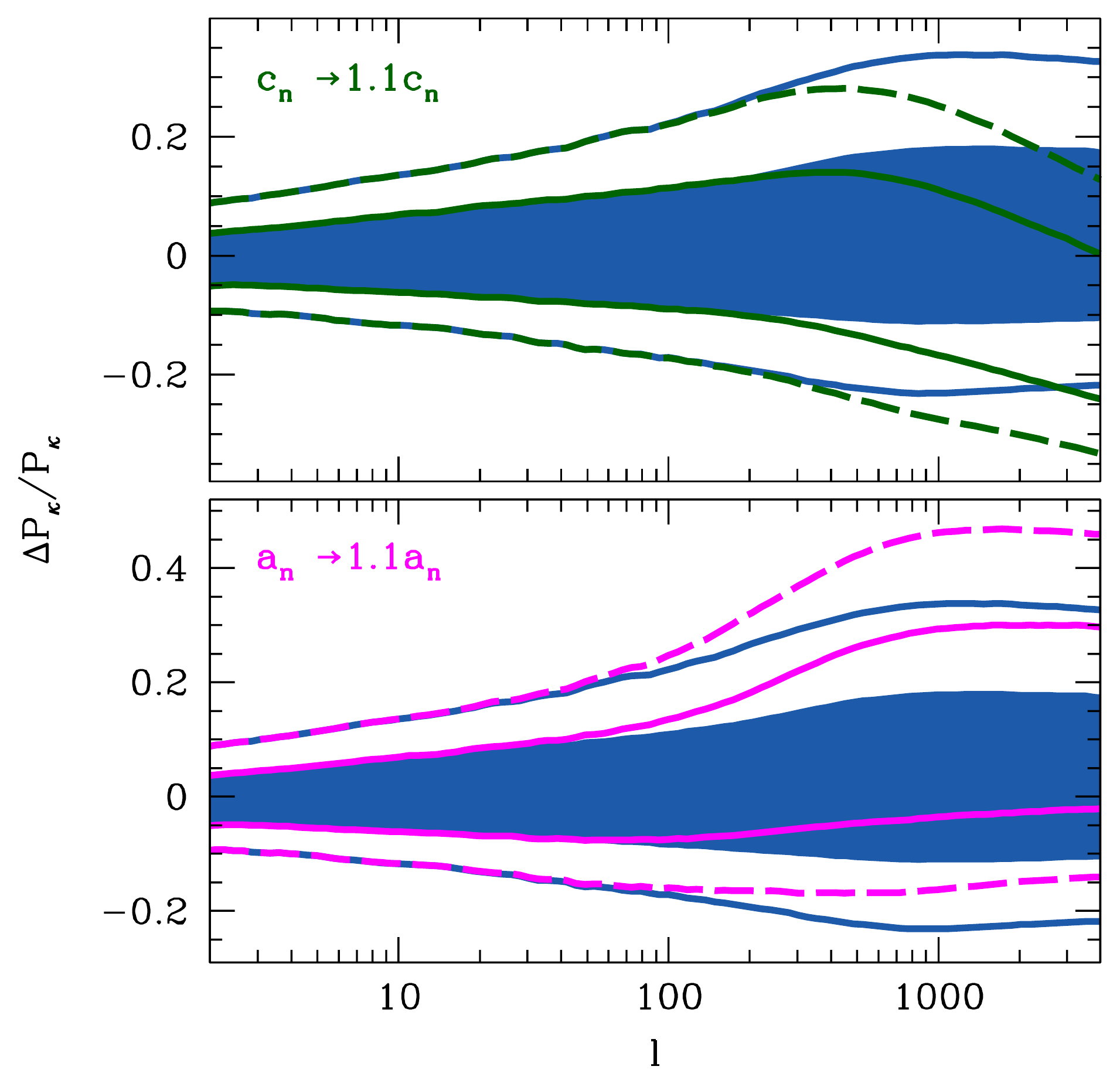}
\caption{Upper panel: Flat $\Lambda$CDM prediction for the CFHTLS shear power spectrum in blue, shown with the prediction for when the parameter $c_n$ is amplified by $10\%$ (green curves). The altered 68\% CL regions are bounded by solid lines and the corresponding altered 95\% CL regions are bounded by dashed lines.
Lower panel: Same as above, but now instead the one-halo term amplitude $a_n$ is amplified by $10\%$ (magenta curves).}
\label{fig:shear_halofit}
\end{figure}

\subsection{Nonlinear Matter Power Spectrum}
\label{sec:nonlinearpk}

We now explore the impact of a few sources of uncertainty in the nonlinear matter power spectrum on our predictions for the shear power spectrum ---
the Halofit fitting function  \cite{Smith:2002dz},
 warm dark matter \cite{Viel:2011bk}, and baryonic physics \cite{Rudd:2007zx,vanDaalen:2011xb}.  We will find it most convenient to parameterize the latter two (physics-based) uncertainties in terms of the former.  For these studies we employ our CFHTLS source redshift distribution and the flat $\Lambda$CDM model class.

Halofit itself has been found to be only accurate at up to the $5$--$10\%$ level in the nonlinear regime for the $\Lambda$CDM cosmological models for which it was designed \cite{Heitmann:2008eq,Heitmann:2009cu,Lawrence:2009uk}, even with the correction factor Eq.~(\ref{halofitcorrection}).  While these inaccuracies are smaller than our current statistical errors in the same regime, they are comparable to the  expected {\it Planck} errors.

Furthermore, we have employed the \lcdm-calibrated Halofit parameters in our predictions for the more general quintessence
class. While we expect that the increased statistical uncertainties in those classes are again currently larger than Halofit errors, this expectation remains largely untested by simulations.
Two of the Halofit parameters in particular describe the characteristic changes in the the nonlinear power spectrum:  $a_n$ controls the amplitude of the one-halo term and $c_n$
describes its shape.     We explore how variations in these two parameters affect our
predictions as a means of quantifying how well Halofit parameters must be calibrated in the \lcdm\ and quintessence $w(z)$ classes.

In Fig.~\ref{fig:shear_halofit},
we show once again our flat $\Lambda$CDM predictions for the CFHTLS shear power spectrum  where in the upper panel we show what happens when $c_n$ is amplified and in the lower panel we show what happens when $a_n$ is amplified. In each case we are only altering the nonlinear matter power spectrum, and thus only the high multipole regime of the shear power spectrum. In the case of amplifying $a_n$, we are simply enhancing the one-halo term. We conclude that $a_n$ and $c_n$ must be calibrated in \lcdm\ to the $\sim 10\%$ level to make use of current statistical predictions for $\ell < 1000$ whereas in quintessence models with EDE and curvature creating large power deficits the tolerances can be  up to double that.

While Halofit was 
constructed to fit N-body simulations with cold dark matter and no baryonic effects,
we find that in particular $c_n$ is a useful parameter
for describing typical deviations from these assumptions.

\begin{figure}[t]
\centering
\includegraphics[width=8cm]{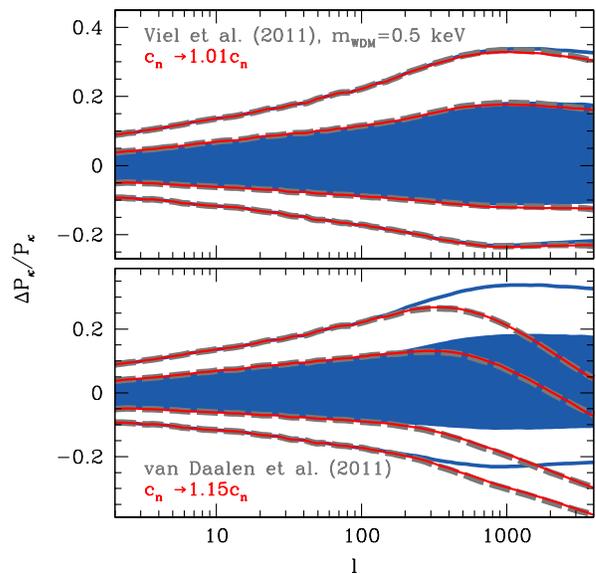}
\caption{Demonstration of our parameterization of WDM (upper panel) and baryonic effects (lower panel) with the Halofit parameter $c_n$, using the CFHTLS source distribution.
Upper panel: Flat $\Lambda$CDM prediction for the shear power spectrum in blue, shown with predictions for $0.5$~keV WDM as in Ref.~\cite{Viel:2011bk} with grey dashed lines and  predictions for $c_n\rightarrow 1.01 c_n$ with red solid lines.
Lower panel: Flat $\Lambda$CDM prediction for the shear power spectrum in blue, shown with predictions for baryons (including AGN feedback) as in Ref.~\cite{vanDaalen:2011xb} with grey dashed lines and  predictions for $c_n\rightarrow 1.15 c_n$ with red solid lines.}
\label{fig:shear_cn}
\end{figure}

Warm dark matter (WDM) reduces the small-scale matter power spectrum in a way that increases with decreasing WDM particle mass. By using high-resolution N-body/hydrodynamic simulations, Ref.~\cite{Viel:2011bk} constructs a fitting function to describe the modified matter power spectrum $P_{\Lambda\rm{WDM}}(k)$ in terms of that of the corresponding $\Lambda$CDM model $P_{\Lambda\rm{CDM}}(k)$,
\be
P_{\Lambda\rm{WDM}}(k)=\left[1+(\alpha k)^{1.8}\right]^{-2/15} P_{\Lambda\rm{CDM}}(k)~,
\ee
where
\be
\alpha(m_{\rm{WDM}},z)=0.0476\left(\frac{1~\rm{keV}}{m_{\rm{WDM}}}\right)^{1.85}\left(\frac{1+z}{2}\right)^{1.3}
\ee
in units of $h^{-1}$~Mpc and $m_{\rm{WDM}}$ is the mass of the warm dark matter particle.   This fitting function is accurate at the $\sim 2\%$ level below $z=3$ and for masses $m_{\rm{WDM}} > 0.5$ keV.  On the other hand, this form assumes a fixed set of \lcdm\ parameters associated
with the simulations and thus effects should be taken as illustrative of the magnitude of the effect.
We find that the WDM effects corresponding to $0.5$ keV dark matter can be mimicked by the Halofit parameter $c_n$
with $c_n\rightarrow 1.01c_n$ (Figure~\ref{fig:shear_cn}, upper panel). 
We thus see that, even for this fairly extreme particle mass, warm dark matter contributes negligibly to the error budget for this particular statistic, in agreement with Ref.~\cite{Viel:2011bk}.   

Baryonic effects, such as radiative heating and cooling, star formation, and supernova and AGN feedback, affect our predictions through their impact on the small-scale matter power spectrum. Given the inherent difficulties in adding baryons to large-scale structure simulations, the degree to which these various processes impact structure growth remains highly uncertain. However we can make a conservative assessment by using the most extreme example, namely the results of Ref.~\cite{vanDaalen:2011xb} which found a significant decrease in power ranging from (at $z=0$) $1\%$ at $k\approx 0.3~h~{\rm Mpc}^{-1}$ to $10\%$ at $k\approx 1~h~{\rm Mpc}^{-1}$ and $30\%$ at $k\approx 10~h~{\rm Mpc}^{-1}$. To find the rough impact on the nonlinear matter power spectrum, we interpolate Fig.~8 of Ref.~\cite{vanDaalen:2011xb}.
Again we find that the Halofit $c_n$ accurately mimics this effect with $c_n\rightarrow 1.15c_n$ corresponding to the simulations of Ref.~\cite{vanDaalen:2011xb} which include AGN feedback (Figure~\ref{fig:shear_cn}, lower panel). A more detailed study of the effect on weak lensing observations is presented in Ref.~\cite{Semboloni:2011fe}.

Thus current uncertainties in baryonic physics at their most extreme are as large as current statistical ranges in the predictions.  Hence the modeling of baryonic effects on structure formation must improve if we are to make use of future constraints above $l=10^3$, as found in Refs.~\cite{Rudd:2007zx,vanDaalen:2011xb}. Note also that, although the particular model we have shown here suppresses scale-scale power, baryons can have the opposite effect of potentially the same magnitude, by increasing small-scale power via cooling \cite{Rudd:2007zx}. 
In the future, one approach to accounting for these systematic errors is to marginalize $a_n$ and $c_n$ as model parameters given
a prior range appropriate to the state-of-the-art simulations including baryonic and other effects.

\begin{figure}[t]
\includegraphics[width=8cm]{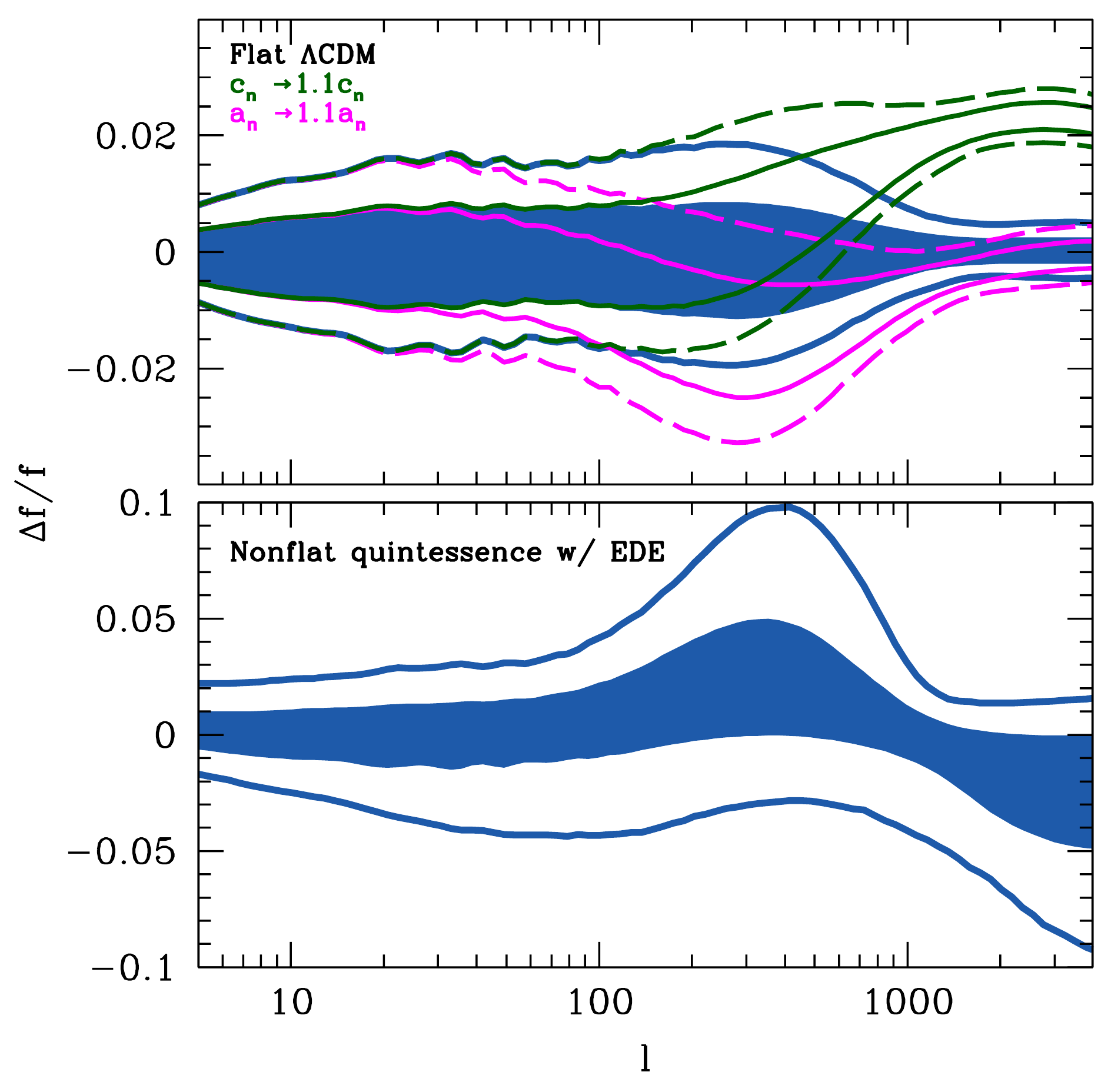}
\caption{The ratio $f(l)$ of the COSMOS and CFHTLS cosmic shear power spectrum predictions for the flat $\Lambda$CDM (upper panel) and nonflat quintessence with EDE (lower panel) model classes, plotted with respect to the ratio of the ML COSMOS and CFHTLS power spectra. Once again shaded regions correspond to 68\% CL and the curves bound the 95\% CL regions. 
The upper panel also shows how the flat $\Lambda$CDM ratio changes when either $c_n$ (upper green contours) or $a_n$ (lower magenta contours) are amplified by $10\%$, where the altered 68\% CL regions are bounded by solid lines and the corresponding altered 95\% CL regions are bounded by dashed lines.}
\label{fig:shear_ratio}
\end{figure}

\subsection{Tomography and Ratio Tests}
\label{sec:tomography}

One means of reducing both the statistical and systematic uncertainty is to employ
tomographic probes that compare the shear at different source redshifts
\cite{Hu:1999ek}.  Uncertainties remaining from the primordial power spectrum 
discussed in \S \ref{sec:linearpk} largely
drop  out and some of the effects of baryonic physics may be ``self-calibrated" through 
jointly determining the concentration-mass relation \cite{Rudd:2007zx,Zentner:2007bn}.

As a simple demonstration and a proxy for tomography, we can define the statistic $f(l)$ (not to be confused with $f(z)$ from Eq.~\ref{eqn:fz}) for a given model class as the ratio of the shear power spectrum $P_{\kappa}(l)$ predicted for COSMOS (median redshift of $1.3$) to that predicted for CFHTLS (median redshift of $0.8$). The resulting prediction is shown in Fig.~\ref{fig:shear_ratio} for our most restrictive (flat $\Lambda$CDM, upper panel) and most general (nonflat quintessence with EDE, lower panel) model classes, where now the $\sim 1\sigma$ uncertainties are reduced to the $1\%$ level for the flat $\Lambda$CDM model. This sharp prediction offers another opportunity to falsify flat $\Lambda$CDM if systematics in the measurement and ratio prediction can be made comparably precise.
In the lower panel of Fig.~\ref{fig:shear_ratio}, the features at small scales are related to the asymmetry of the quintessence growth predictions and the difference in the nonlinear scales for the source distributions of COSMOS ($l\gtrsim 200$) and CFHTLS ($l\gtrsim 500$).

However, present uncertainties in the nonlinear power spectrum arising from Halofit render predictions in the nonlinear regime unreliable. 
In the upper panel of Fig.~\ref{fig:shear_ratio} we show how the uncertainties in $c_n$ (green, upper curves) and $a_n$ (magenta, lower curves), respectively, affect the flat \lcdm\ predictions for $f(l)$. These uncertainties affect the quintessence predictions similarly. We see that $\sim 10\%$ level calibration of these Halofit parameters is not sufficient to exploit the sub-percent level \lcdm\ predictions in the nonlinear regime.  On the other hand, observed deviations in $f(l)$ of $\gtrsim 3\%$ would falsify \lcdm\  and  $\gtrsim 10\%$ our most general quintessence class even considering these uncertainties. 

\subsection{Supernova Light-Curve Fitting}
\label{sec:sn}

The analyses of the datasets outlined in \S \ref{sec:data} contribute further systematic uncertainties in our predictions, the largest of which arise from the fitting of SN light-curves. The two most widely used light-curve fitters are SALT2 and MLCS2k2. Assuming a flat cosmological model with a constant dark energy equation of state $w$, the SALT2 and MLCS2k2 methods yield  $(\Omega_{\rm m},w)=(0.26 \pm 0.03, -0.96 \pm 0.13)$ and $(\Omega_{\rm m},w)=(0.31 \pm 0.03, -0.76 \pm 0.13)$, respectively \cite{SDSS_SN}. Thus these two methods lead to a discrepancy in the dark energy equation of state of $\Delta w \sim 0.2$. This might be due to a possible mismatch in the UV spectra between low-redshift and intermediate-redshift SNe \cite{Foley:2010mm} and efforts are underway to reduce the resulting error. We have chosen here to use SN data analyzed using the SALT2 technique, but we now briefly discuss what happens when we instead use the MLCS2k2 data.\footnote{Note that Ref.~\cite{Mortonson:2010mj} made the opposite choice, using the MLCS2k2 SN constraints for their main results.}

\begin{figure}[t]
\includegraphics[width=8cm]{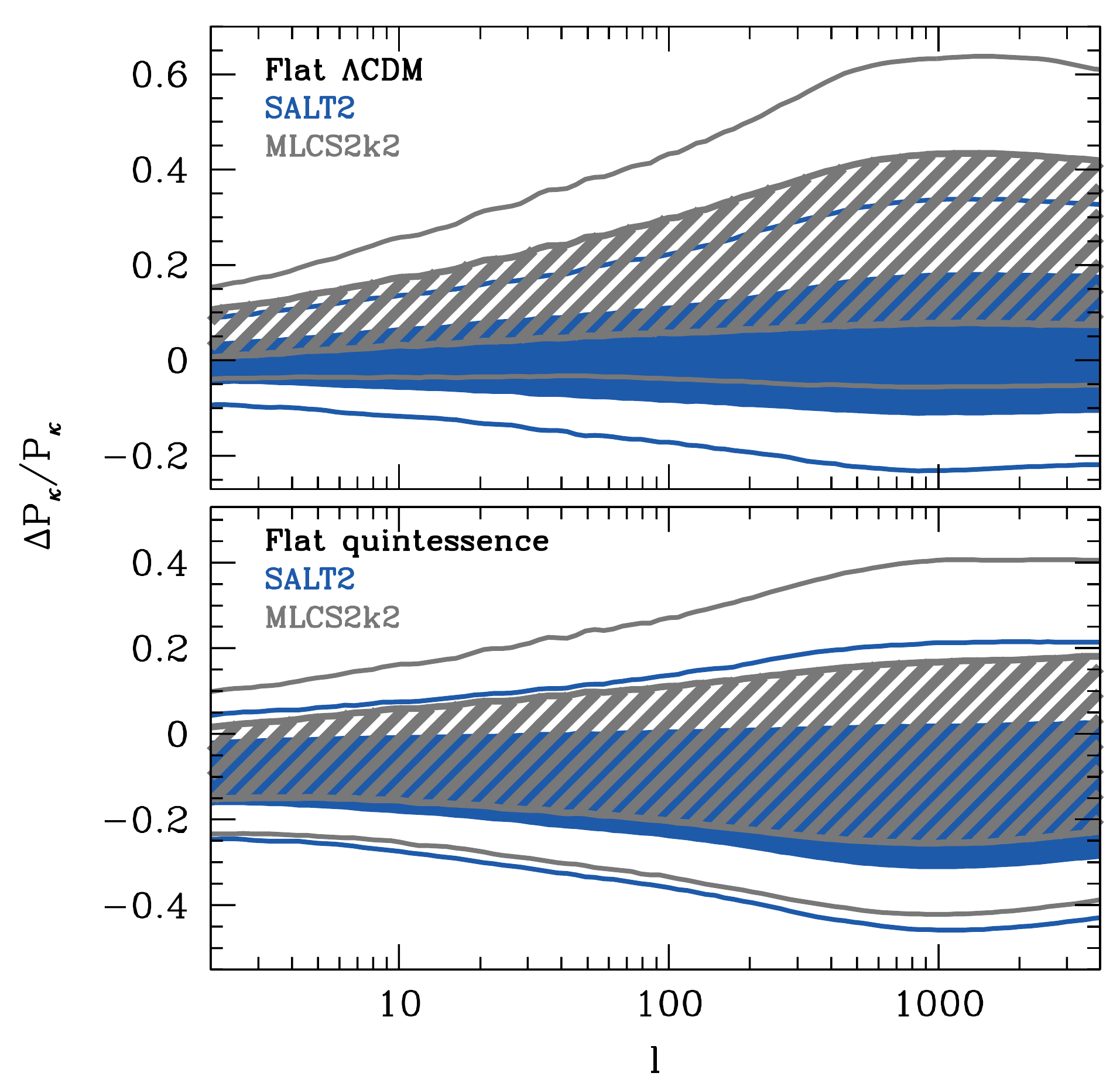}
\caption{Shear power spectrum predictions for the CFHTLS source redshift distribution, where we show the bias resulting from the choice of SN light curve fitter. 
Upper panel: Flat $\Lambda$CDM predictions for SALT2 (blue solid) and MLCS2k2 (grey hatched), plotted relative to the ML SALT2 flat $\Lambda$CDM result. Once again the shaded regions correspond to 68\% CL and the curves bound the 95\% CL regions. 
Lower panel: Flat quintessence predictions for SALT2 (blue solid) and MLCS2k2 (grey hatched), also plotted relative to the ML SALT2 flat $\Lambda$CDM result.}
\label{fig:shear_ms}
\end{figure}

Even for the flat $\Lambda$CDM model class, the differences between SALT2 and MLCS2k2 estimates of $\Omega_{\rm m}$ and $w$ significantly alter our predictions for the shear power spectrum, as can be seen in the upper panel of Fig.~\ref{fig:shear_ms}. In the lower panel we show the comparison for flat quintessence; in both cases, the CL regions shift by $\sim 0.5$--$1\sigma$. This shift to higher shear with MLCS2k2 is driven mainly by the preference for higher $\Omega_{\rm m}$ when using this method; this higher $\Omega_{\rm m}$ also increases the present day matter power spectrum normalization for a fixed amplitude of initial curvature fluctuations $A_s$. The corresponding effect for cluster abundance was found in Ref.~\cite{Mortonson:2010mj}.
With improvements in the calibration of SN data and modeling of the 
effects of dust extinction, it is likely that these systematics can be reduced
so that they will not be a limiting uncertainty for future predictions of 
shear statistics.

\section{Discussion}
\label{sec:discussion}

In this paper we have provided robust statistical predictions for weak lensing observables in a variety of cosmological contexts. Given existing local distance measures of $H_0$, the Type Ia SNe distance-redshift relation, BAO distance measures, and the CMB temperature and polarization power spectra, we have constrained the expected cosmic shear power spectrum in the $\Lambda$CDM model. We then generalized this model class to show how the predictions widen when curvature, EDE, and a generalized dark energy equation of state $w(z)$ are allowed. We further showed how these predictions are affected by uncertainties in the nonlinear matter power spectrum from warm dark matter and baryons, to find that the former is negligible whereas the latter is a comparable source of uncertainty to current statistical errors beyond $l\sim 10^{3}$.  In the near future baryonic effects will likely become the dominant source of error unless modeling improves.

In a similar fashion as in the clusters study of Ref.~\cite{Mortonson:2010mj}, we find that any observation that claims to rule out the $\Lambda$CDM model via a shear excess generically also rules out the entire quintessence paradigm, as the extra freedom allowed by the free function $w(z)\ge -1$ can only serve to reduce the relative growth if we normalize to the CMB and constrain the distance-redshift relation to the CMB and SNe at opposite ends of the
expansion history.
Adding EDE, warm dark matter, massive neutrinos, or baryonic AGN feedback tends to exacerbate the reduction, as they too only reduce growth and therefore suppress the cosmic shear power spectrum. Therefore a measured shear excess would require updating our models to include significant baryonic cooling, primordial non-Gaussianity, non-smooth dark energy, 
phantom dark energy,  an interacting dark sector, or even a modification to the gravitational
force law.

In our analysis we have focused only on those uncertainties pertaining to the predictions for the shear statistics presented here, as opposed to uncertainties in the measurement of
the shear itself.   The latter becomes important when comparing the data to these predictions. 
For example in comparing our predictions to current data in Fig.~\ref{fig:xip_comp}, we have
employed the statistical and systematic error estimates in the literature. Uncertainties relating to shape measurement, photometric redshifts, and intrinsic alignments feed into the data error covariance.   Accounting for these effects, the flat \lcdm\ model is consistent with the
current data.

Future imaging surveys, both from the ground (e.g.~DES and LSST) and from space ({\it Euclid} and/or {\it WFIRST}) will provide the necessary number of galaxy shape measurements to enable precise tests of dark energy model classes, as long as the various shear systematics can be either eliminated or understood sufficiently well. Moreover, improved 
CMB constraints from {\it Planck} will soon reduce the uncertainties in the 
cosmic shear predictions by a factor of a few, further enhancing the ability 
of weak lensing observations to detect deviations from the concordance cosmological model but also requiring more stringent control on astrophysical systematic errors in both the predictions and the measurements.

~~~~~~~~~~~~~~~~~~~~~~~~~~~~~~

\begin{acknowledgments}

We thank Chris Hirata, Richard Massey, Catherine Heymans, and Tim Schrabback for assistance regarding weak lensing survey parameters and results and Scott Dodelson, Eduardo Rozo, Elise Jennings, David Weinberg, and Mark Wyman for useful conversations. RAV and WH acknowledge the support of the Kavli Institute for Cosmological Physics at the University of Chicago through grants NSF PHY-0114422 and NSF PHY-0551142 and an endowment from the Kavli Foundation and its founder Fred Kavli. MJM and TE acknowledge support from CCAPP at the Ohio State University. WH acknowledges additional support from
DOE contract DE-FG02-90ER-40560 and the Packard Foundation.

\end{acknowledgments}

\bibliography{pcshear}

\end{document}